\documentclass[referee]{mn2e} 
\usepackage{amssymb} 
\usepackage{times}
\input{psfig.sty} 
\begin{document}   
   
\title{Relativistic $r$-modes and Shear viscosity: regularizing the 
continuous spectrum.} 
  
\author[J.A. Pons {\it et al.}] 
{J.A. Pons $^1$, L. Gualtieri$^2$, J.A. Miralles $^1$, V. Ferrari $^2$ 
\\   
$^1$ Departament de F\'{\i}sica Aplicada, Universitat d'Alacant, 
Apartat de correus 99, 03080 Alacant, Spain 
\\ 
$^2$ Dipartimento di Fisica ``G.Marconi",   
 Universit\` a di Roma ``La Sapienza"\\   
and Sezione INFN  ROMA1, piazzale Aldo  Moro   
2, I-00185 Roma, Italy  } 
   
\maketitle   
  
\begin{abstract}   
Within a fully relativistic framework, we derive and solve numerically 
the perturbation equations of relativistic stars, including the 
stresses produced by a non-vanishing shear viscosity in the 
stress-energy tensor. With this approach, the real and imaginary parts 
of the frequency of the modes are consistently obtained.
We find that, approaching the inviscid limit from the 
finite viscosity case, the continuous spectrum is regularized and we 
can calculate the quasi-normal modes for stellar models that do not 
admit solutions at first order in perturbation theory when the 
coupling between the polar and axial perturbations is neglected. 
The viscous damping time is found to agree within factor 2 
with the usual estimate obtained by using the eigenfunctions of the 
inviscid limit and some approximation for the energy dissipation 
integrals. We find that the frequencies and viscous damping times for 
relativistic $r-$modes lie between the Newtonian and Cowling results. 
We compare the results obtained with homogeneous, polytropic and 
realistic equations of state and find that the frequencies
depend only on the rotation rate and on the compactness parameter $M/R$, 
being almost independent of the equation of state. 
Our numerical results for realistic neutron stars give viscous damping
times with the same dependence on mass and radius as previously estimated,
but systematically larger of about 60\%.
  
\end{abstract}   

  
\section{Introduction}   
  
Since the discovery of the $r-$mode instability (Andersson, 1998), the 
study of their astrophysical relevance has become a continuously 
growing field.  In particular, the gravitational radiation from the 
$r-$ mode instability has been proposed to be the reason why  
all observed young neutron stars have rotational frequency much 
smaller than the Keplerian frequency (Andersson {\em et al.}, 1999; 
Lindblom {\em at al.}, 1998), to provide a way to identify strange 
stars as persistent sources of gravitational waves (Andersson {\em et 
al.}, 2002), or to be the reason why accreting neutron stars in low 
mass $X$-ray binaries rotate in a narrow range of frequencies 
(Bildsten, 1998; Wagoner, 2002). A comprehensive review on 
gravitational waves from instabilities in relativistic stars, 
including the $r-$mode instability, can be found in Andersson (2003). 
  
Several issues concerning the $r-$modes have been object of 
controversy, among which we must mention the existence of the 
continuous spectrum, and the way in which viscosity can stabilize the 
stars.  It was soon pointed out (Kojima, 1998; Ruoff \& Kokkotas, 
2001) that, working at first order in perturbation theory and 
neglecting the coupling between different harmonics and between the 
polar and axial perturbations, leads to the existence of a continuous spectrum 
and makes doubtful the existence of the modes. As shown later, going 
beyond the low frequency limit and the Cowling approximation does not 
remove the continuous spectrum (Ruoff \& Kokkotas, 2002). However, 
hydrodynamical numerical simulations (Gressman {\em et al.}, 2002), or 
evolutionary descriptions taking into account the polar/axial coupling 
(Villain \& Bonazzola, 2002), seem to indicate that $r-$modes do 
exist, so that the continuous spectrum may be interpreted as an 
artifact due to an inconsistency of the perturbative expansion when 
$\sigma\sim\Omega$.  With this motivation, more sophisticated methods 
to solve the eigenvalue problem in a general inviscid case have been 
developed (Lockitch {\em et al.}, 2001; Lockitch {\em et al.}, 2003; 
Lockitch {\em et al.}, 2004).  On the other hand, in any realistic 
situation viscosity is actually present. An interesting question to 
address is then if the inclusion of viscosity in the game does affect 
the existence of the continuous spectrum, as suggested for example in 
Lockitch {\em et al.} (2004). 
It is believed that bulk/shear viscosity limit the instability at 
high/low temperatures, respectively.  But the situation is 
complicated because the effects of superfluidity in 
the inner core, hyperon viscosity, or the core-crust shear layer are 
uncertain (Andersson {\em et al.}, 2004; Glampedakis \& 
Andersson, 2004; Lindblom \& Owen, 2002).  

To our knowledge, in
previous works the effect of viscosity in damping the $r-$modes
has been studied in the following way.  The star is assumed to be
composed of a perfect fluid and  the eigenvalues and 
eigenfunctions of the modes are computed by solving the adiabatic, perturbed 
Newtonian or Einstein's equations. The viscous damping time is then computed
by evaluating some suitably defined  integrals that express the energy content 
of the modes. This approach has been used, for example,
in the seminal work about viscosity on neutron 
star oscillations by Cutler \& Lindblom, (1987), and it
is justified by the fact that viscosity is indeed very small.
However, even a very small amount of viscosity may be crucial to 
change the role of the $r$-mode instability in astrophysical processes, 
therefore this limit deserves to be studied more accurately.  In this work, 
we introduce the shear viscosity in the stress-energy tensor from 
the beginning, and we evaluate consistently both the 
frequency and the damping time of the mode. By this approach, one 
could determine more precisely the window of instability, for a given 
stellar model, provided that the microphysical input (EOS, viscosity) 
is known.  A similar self-consistent inclusion of the heat transfer 
corrections to the polar perturbations has recently been done 
(Gualtieri {\em et al.}, 2004). An interesting results of this work is 
that shear viscosity, even if very small, regularizes the continuous spectrum.
  
The paper is organized as follows. In Sec. 2 we derive the equations 
of stellar perturbation with shear viscosity. In Sec. 3 we discuss 
the numerical integration of the perturbed equations and the boundary 
conditions.  In Sec. 4 we examine the results for the different 
stellar models (homogeneous, polytropic, realistic) and 
in Sec. 5 we draw the conclusions and discuss the domain of 
applicability of our approach and future extensions. In the Appendix, 
an analytic solution for Newtonian $r-$modes with viscosity is 
derived.

\section{The perturbed equations with shear viscosity}   
  
We consider a star rotating uniformly with angular velocity $\Omega$. 
At first order in $\Omega$ (or, more precisely, at first order in the 
rotational parameter $\epsilon\equiv\Omega/\Omega_N$ with 
$\Omega_N\equiv\sqrt{2M/R^3}$), the stationary background is described 
by the metric (Hartle, 1967; Hartle \& Thorne, 1968) 
\begin{equation} 
ds^2=g^{(0)}_{\mu\nu}dx^\mu dx^\nu= 
-e^{\nu(r)}dt^2+e^{\lambda(r)}dr^2+r^2(d\vartheta^2+\sin^2\vartheta 
d\varphi^2) -2r^2\omega(r)\sin^2\vartheta d\varphi dt\,,\label{metric} 
\end{equation}  
where $\omega(r)$ represents the dragging of the inertial frames.  It 
corresponds to the angular velocity of a local ZAMO (zero angular 
momentum observer), with respect to an observer at rest at infinity. 
The $4$--velocity of the fluid is simply 
$u^{(0)\mu}=(e^{-\nu/2},0,0,\Omega e^{-\nu/2})$, and the stress-energy 
tensor is 
\begin{equation}   
T^{(0)}_{\mu\nu}=(\rho+p)u^{(0)}_{\mu}u^{(0)}_{\nu}+pg^{(0)}_{\mu\nu}\,.  
\end{equation}  
We assume that viscosity does not affect the stationary axisymmetric 
background, because the shear tensor and the expansion do vanish 
there. Therefore, in this regime the Einstein field equations reduce 
to the standard TOV (Tolman-Oppenheimer-Volkoff) equations plus a 
supplementary equation for the frame dragging $\omega(r)$: 
\begin{equation}   
\bar\omega_{,rr}-\left(4\pi(\rho+p){e^\lambda} 
r-\frac{4}{r}\right)\bar\omega_{,r} 
-16\pi(\rho+p){e^\lambda}\bar\omega=0\label{hartleeq} 
\end{equation}  
where $\bar\omega(r)\equiv\Omega-\omega(r)$.  
  
\subsection{Perturbations with axial parity}  
In this paper we focus on purely axial perturbations of the metric 
(\ref{metric}).  The perturbed metric, expanded in spherical harmonics 
and in the Regge-Wheeler gauge, depends only on the axial vector 
harmonics 
\begin{equation}   
S^{lm}_a=\left(-\frac{1}{\sin\vartheta}Y^{lm}_{,\varphi},  
\sin\vartheta Y^{lm}_{,\vartheta}\right)  
\end{equation}  
where $a=\vartheta,\varphi$ and $Y^{lm}(\vartheta,\varphi)$ are the 
scalar spherical harmonics (in the following we omit the harmonic 
indexes $lm$).  The Einstein equations can be shown to reduce to a set 
of equations for two metric variables ($h_0, h_1$) and for the axial 
component of the fluid velocity (Kojima, 1992), whose harmonic 
expansions read: 
\begin{eqnarray}  
h_{0a}(t,r,\vartheta,\varphi)&=& 
h_0(r,\sigma)S_a(\vartheta,\varphi)e^{-{\rm i}\sigma t}\nonumber\\ 
h_{1a}(t,r,\vartheta,\varphi)&=& 
h_1(r,\sigma)S_a(\vartheta,\varphi)e^{-{\rm i}\sigma t}\nonumber\\  
\delta u^a(t,r,\vartheta,\varphi)&=& 
\frac{e^{-\nu/2}}{r^2}Z(r,\sigma)\gamma^{ab} S_b(\vartheta,\varphi)
e^{-{\rm i}\sigma t}. 
\label{harmexp} 
\end{eqnarray}  
Here, $\gamma_{ab}=$diag$(1,\sin^2\vartheta)$ is the metric on the 
two-sphere, $\sigma$ is a (complex) frequency, $\delta u^r=0$, and 
$\delta u^t$ can be obtained imposing $u^\mu u_\mu=-1$. The axial 
perturbations are then described by only three degrees of freedom: 
two, $h_0,h_1$, for the metric, and one, $Z$, for the fluid.  Notice 
that for the metric variables we follow the convention used in Kojima 
(1998) and Ruoff \& Kokkotas (2002), but we use a different fluid 
variable.  Our variable $Z$ is related to the $u_3$ in Ruoff \& 
Kokkotas (2002) (indicated with $RK$) by 
\begin{equation}   
Z=e^{(\nu-\lambda)/2}u_3^{RK}-h_0  
\end{equation}  
and our functions $\lambda,\nu$ are related to those used in Ruoff \& 
Kokkotas (2002) by $\lambda = 2\lambda^{RK}\,,~\nu= 
2\nu^{RK}\,$. In order to simplify the form of Einstein's 
equations we define the following variables 
\begin{eqnarray}  
K&=&e^{(\lambda-\nu)/2}h_0\label{defK}\\  
V&=&-{\rm i} e^{(\nu-\lambda)/2}h_1\label{defV}\\  
H&=&{e^{-\lambda}}\left[K'+\left(\frac{\nu'-\lambda'}{2}-\frac{2}{r}\right)K  
-(\sigma-m\omega)e^{\lambda-\nu}V\right]\label{defH}  
\end{eqnarray}  
which are related to those of Ruoff \& Kokkotas, (2002) by  
\begin{eqnarray}  
K&=&K_6^{RK}\nonumber\\  
V&=&-{\rm i} V_4^{RK}\nonumber\\  
H&=&K_3^{RK}\,.  
\end{eqnarray}  
Notice that we have introduced an auxiliary variable, $H$, in order to 
have, at least in the perfect fluid case, a system of first order 
differential equations. 
  
\subsection{Einstein's equations}  
Taking into account viscosity, the stress-energy tensor  
has the form (Misner {\em et al.}, 1973)  
\begin{equation}   
T_{\mu\nu}=(\rho+p)u_{\mu}u_{\nu}+pg_{\mu\nu}-2\eta\sigma_{\mu\nu}  
-2\zeta g_{\mu\nu}u^{\alpha}_{~;\alpha}\,.  
\end{equation}  
Here $\eta,\,\zeta$ are the shear and bulk viscosity coefficients (see 
Cutler \& Lindblom, (1987) for a discussion on the meaning of these 
coefficients), and 
\begin{equation}   
\sigma_{\mu\nu}\equiv\frac{1}{2}g^{\rho\lambda}\left(u_{\mu;\rho}  
P_{\nu\lambda}+u_{\nu;\rho}P_{\mu\lambda}\right)-\frac{1}{3}u^{\rho}_{~;\rho}  
P_{\mu\nu}  
\end{equation}  
with $P_{\mu\nu}$ being the projector onto the subspace orthogonal to 
$u_{\mu}$ 
\begin{equation}   
P_{\mu\nu}\equiv g_{\mu\nu}+u_\mu u_\nu\,.  
\end{equation}  
In the interior of the star, after expanding the perturbed tensors in 
spherical harmonics, Einstein's equations perturbed at first order are: 
\begin{eqnarray}  
K'&=&-\left(\frac{\nu'-\lambda'}{2}-\frac{2}{r}\right)K  
+(\sigma-m\omega)e^{\lambda-\nu} V+e^{\lambda}H\label{eqKint2}\\  
V'&=&-(\sigma-m\omega)K+\frac{mr^2}{\Lambda}\left[\omega'H-  
16\pi(\rho+p)\bar\omega\left(K+Ze^{(\lambda-\nu)/2}\right)\right]\nonumber\\  
&&+16\pi{\rm i}\eta\left[e^{\lambda/2}Z-\frac{m}{\Lambda}\sigma\bar\omega r^2  
\left(Ze^{\lambda/2-\nu}+Ke^{-\nu/2}\right)\right]\nonumber\\  
&&\label{eqVint2}\\  
H'&=&-\frac{2}{r}H+\frac{\Lambda-2}{r^2}K+16\pi(\rho+p)\left[Ze^{(\lambda-\nu)/2}+  
K\right]-2\frac{m}{\Lambda}e^{-\nu}\omega'V\nonumber\\  
&&-16\pi{\rm i}\eta m\Omega\frac{\Lambda-2}{\Lambda}e^{\lambda/2-\nu}Z  
\label{eqHint2d}\\  
&&\nonumber\\  
- (\sigma-m\omega) H&=& \frac{\Lambda-2}{r^2}V  
+\frac{2m\omega'}{\Lambda}e^{-\lambda}K  
-16\pi{\rm i}\eta X  
\label{eqHint2}\\  
X&=& e^{-\lambda/2}\left(Z'-\frac{2}{r}Z\right)  
+ (\sigma-m\Omega) e^{-\nu/2}V  
\label{Xint}  
\\  
\left(\sigma-m\Omega+\frac{2m\bar\omega}{\Lambda}\right)Z &=&  
-(\sigma-m\Omega) e^{(\nu-\lambda)/2}K  
~+ ~ {\rm i}\frac{e^{(\nu-\lambda)/2}}{(\rho+p)r^2} \times  
\nonumber\\  
&\times&\left\{  
\left(r^2 \eta X \right)'   
- \eta {(\Lambda-2)} \left[ e^{\lambda/2} Z -  
\frac{r^2 m\sigma\bar\omega}{\Lambda} \left( e^{\lambda/2-\nu} Z  
+ e^{-\nu/2} K \right) \right] \right\}  
\nonumber\\  
\label{eqZint2}  
\end{eqnarray}  
where $\Lambda\equiv l(l+1)$ and we have introduced the new variable $X$ 
(Eq. (\ref{Xint})).
We note that this quantity comes directly from the expansion in
axial vector harmonics of the shear tensor components 
$\sigma_{r\vartheta}$, $\sigma_{r\varphi}$.
Having written the equations, we would like to make some relevant 
remarks. 
\begin{itemize}  
\item We have neglected the coupling between axial perturbations with 
harmonic index $l$ and polar perturbations with harmonic indexes $l\pm 
1$, as in Ruoff \& Kokkotas (2002) and Kojima (1993).  The reason is that 
in this paper we focus on the effect of viscosity on $r$-mode 
oscillations, and we want to separate different effects. We point out 
that the coupling terms between polar and axial perturbations affects
the shape of the continuous spectrum (Ruoff {\em et al.}, 2002). 
However, in this paper we show that, having neglected this coupling, 
viscosity removes the continuous spectrum. 
\item Under the approximation described above, bulk viscosity is not 
coupled to axial perturbations, because it enters into the equations 
only through the axial-polar $l\leftrightarrow l\pm 1$ 
couplings. Thus, we will only study the effects of shear viscosity, 
leaving the investigation of the effects of bulk viscosity for future 
work. 
\item In the perfect fluid case $\eta=0$, we have three first order 
differential equations, (\ref{eqKint2}--\ref{eqHint2d}), and three
algebraic relations, (\ref{eqHint2}--\ref{eqZint2}).  Notice that $H$ 
can be given, alternatively, by the differential equation 
(\ref{eqHint2d}) or by the algebraic relation (\ref{eqHint2}). While 
the inviscid limit of equations (\ref{eqKint2}--\ref{eqVint2}) and 
(\ref{eqHint2}--\ref{eqZint2}) has been derived before (e.g. Ruoff \& 
Kokkotas, (2002)), the differential equation for $H$ (\ref{eqHint2d}) 
has never been derived so far; we have checked numerically that it is 
equivalent to determine $H$ by (\ref{eqHint2}) or by integration of 
(\ref{eqHint2d}). 
\item In the perfect fluid case $\eta=0$ the equation (\ref{eqZint2}) 
for $Z$ is a simple algebraic relation, and $Z$ can be found by 
dividing by 
$\left(\sigma-m\Omega+\frac{2m\bar\omega}{\Lambda}\right)$; this is 
the origin of the continuous spectrum which appears when this term 
vanishes.  When viscosity is present, extra terms do appear involving 
the second derivative of $Z$ so that, in order to solve our equations 
in the frequency domain, we prefer to recast 
Eqs. (\ref{Xint}--\ref{eqZint2}) as 
\begin{eqnarray}  
Z'&=& \frac{2}{r} Z + e^{\lambda/2} {X} - (\sigma-m\Omega)  
e^{(\lambda-\nu)/2}V \label{Zp}\\  
X'&=& -\left(\frac{2}{r}+\frac{\eta'}{\eta}\right) X +  
\frac{(\Lambda-2)}{r^2} \left[ e^{\lambda/2} Z -  
\frac{r^2 m\sigma\bar\omega}{\Lambda} \left( e^{\lambda/2-\nu} Z  
+ e^{-\nu/2} K \right) \right]  
\nonumber\\  
&&-{\rm i}\frac{(\rho+p)}{\eta}e^{(\lambda-\nu)/2}\left[  
\left(\sigma-m\Omega+\frac{2m\bar\omega}{\Lambda}\right)Z +  
(\sigma-m\Omega) e^{(\nu-\lambda)/2}K\right].
\label{Xp}  
\end{eqnarray}  
 
Because of numerical problems associated to the $1/\eta$ term, we 
cannot integrate this system of equations for 
$\eta\lesssim10^{-6}$ km$^{-1}$. \footnote{Usually the typical 
viscosity at $T=10^7K$ and $\rho=10^{15}$g/cm$^3$ is $\eta\sim 
10^{23}~{\rm g/cm/s}=2.4\cdot10^{-11}$km$^{-1}$.} 
\end{itemize}  
  
In the exterior of the star, the stress-energy tensor vanishes and the 
equations for the metric perturbations are 
\begin{eqnarray}  
K'&=&\left(\lambda'+\frac{2}{r}\right)K  
+(\sigma-m\omega)e^{2\lambda}V+e^{\lambda}H\label{eqKext}\\  
V'&=&-(\sigma-m\omega)K+\frac{mr^2}{\Lambda}\omega'H\label{eqVext}\\  
H'&=&-\frac{2}{r}H+\frac{\Lambda-2}{r^2}K-  
2\frac{m}{\Lambda}{e^\lambda}\omega'V\label{eqHextd}\\  
H&=&-\frac{1}{\sigma-m\omega}\left[\frac{\Lambda-2}{r^2}V  
+\frac{2m\omega'}{\Lambda}e^{-\lambda}K\right]\label{eqHext}  
\end{eqnarray}  
where we can alternatively use (\ref{eqHextd}) or (\ref{eqHext})   
to determine $H$.  
  
\subsection{Cowling approximation and Newtonian limit}  
Since in Section \ref{results} we shall make a detailed comparison 
with previous works, it is useful to write the equations in the 
Cowling approximation and in the Newtonian limit.  The equations in 
the relativistic Cowling approximation can easily be obtained. By 
neglecting all metric perturbations, and considering as dynamical 
equations of the system $\delta T^{\mu\nu}_{~;\nu}=0$, we end up with 
a single second order differential equation for the fluid variable 
$Z$, which we write as a system of two first order equations: 
\begin{eqnarray}  
Z'&=& \frac{2}{r} Z + e^{\lambda/2} X  
\\  
X'&=& -\left( \frac{2}{r}+ \frac{\eta'}{\eta} \right) X +  
\frac{(\Lambda-2)}{r^2} \left[ 1 -  
\frac{r^2 m\sigma\bar\omega e^{-\nu}}{\Lambda}\right] e^{\lambda/2} Z  
\nonumber\\  
&&-{\rm i}\frac{(\rho+p)}{\eta}e^{(\lambda-\nu)/2}   
\left(\sigma-m\Omega+\frac{2m\bar\omega}{\Lambda}\right) Z~.  
\label{eqscowling}  
\end{eqnarray}  
Notice that in the limit $\eta \rightarrow 0$, it reduces to   
\begin{eqnarray}  
\left(\sigma-m\Omega+\frac{2m\bar\omega}{\Lambda}\right)Z = 0  
\end{eqnarray}  
showing the well known problem of the continuous spectrum due to the 
dragging of the inertial frames. The Newtonian limit can be obtained 
by taking $e^{\lambda}=1, e^{\nu}=1$, pushing the speed of light to 
infinity (so that in our units $p/\rho\rightarrow 0$) and neglecting 
the relativistic corrections ($\bar\omega=\Omega$). The resulting 
equations are: 
\begin{eqnarray}  
Z'&=& \frac{2}{r} Z + X  
\label{eqsnewt0}\\  
X'&=& -\left( \frac{2}{r}+ \frac{\eta'}{\eta} \right) X +  
\frac{(\Lambda-2)}{r^2} \left[ 1 -  
\frac{r^2 m\sigma\Omega }{\Lambda}\right] Z  
-{\rm i} \frac{\rho}{\eta}  
\left(\sigma-\frac{\Lambda-2}{\Lambda}m\Omega\right) Z  
\label{eqsnewt}  
\end{eqnarray}  
with the very well known inviscid   
limit $\sigma=\frac{(\Lambda-2)}{\Lambda}m \Omega$.  
  
As shown in the Appendix, equations (\ref{eqsnewt0}--\ref{eqsnewt}) 
admit an analytical solution for models with constant $\rho$ and $\eta$. 
Furthermore, with a suitable change of 
variable, they can be integrated numerically for any value of $\eta$. 
\section{Numerical implementation}   
The system of equations (\ref{eqKint2}--\ref{eqHint2d}), (\ref{Zp}), 
(\ref{Xp}) coupled to the TOV equations and Hartle's equation 
(\ref{hartleeq}) is solved by means of an adaptive step, third order 
Runge-Kutta integration scheme. 
After integrating the equations in the interior of the star, we 
integrate the exterior equations (\ref{eqKext}--\ref{eqHext}) backward 
starting with an expansion in terms of the purely outgoing solutions 
at a large radius, as explained in Sec. \ref{outgoingcond}.  By 
evaluating the Wronskian of the interior and exterior solutions, we 
can obtain the quasi-normal mode frequency by monitoring when the 
modulus of the Wronskian vanishes. We have implemented a 
Newton-Raphson scheme to find the correct frequency at which the 
Wronskian vanishes starting with an initial guess.  In the inviscid 
case, there is only one independent solution that is regular at the 
origin; in the viscous case there are two independent solutions 
satisfying the boundary conditions at the center, as we show 
below. Thus, an additional boundary condition will be required at the 
star's surface. 
  
\subsection{Boundary conditions at the center}  
Expanding equations (\ref{eqKint2}--\ref{eqZint2}) near   
$r=0$, we find the following behaviour: 
\begin{eqnarray}  
K&=&r^{l+1}\nonumber\\  
V&=&Ar^{l+2}\nonumber\\  
H&=&(l-1)r^l\nonumber\\  
Z&=&Cr^{l+1}+Dr^{l+3}\label{center}  
\end{eqnarray}  
where  
\begin{eqnarray}  
A&=&\frac{1}{l+2}\left[-(\sigma-m\omega_c)+16\pi{\rm i}\eta 
C\right]\label{defAA}~,\\  
C&=&\frac{-(\sigma-m\Omega)e^{\nu_c/2}+ \frac{{\rm i}\eta}{\rho_c+p_c} 
{\cal A} } {\left(\sigma-m\Omega+\frac{2m\bar\omega_c}{\Lambda} 
\right) +\frac{{\rm i}\eta}{\rho_c+p_c}{\cal B} }~, 
\end{eqnarray}  
and 
\begin{eqnarray}  
{\cal A}& =& \left[-\frac{l+4}{l+2}(\sigma-m\Omega)(\sigma-m\omega_c)+  
m\sigma\bar\omega_c\frac{\Lambda-2}{\Lambda}  
+2e^{\nu_c/2}(2l+3)D\right]~,  
\nonumber \\  
{\cal B}& =& e^{\nu_c/2}\left[(l-1)(2l+5)  
\frac{8}{3}\pi\rho_c-16\pi{\rm i}\eta\frac{l+4}{l+2}e^{-\nu_c/2}  
(\sigma\!-\!m\Omega)  
-2m\sigma\bar\omega_ce^{-\nu_c}\frac{\Lambda-2}{\Lambda}\right]~.  
\nonumber  
\end{eqnarray}  
This last relation couples the constant $D$ to the others. The reason 
why we need the term proportional to $r^{l+3}$ in the expansion of $Z$ 
is that, keeping only the leading term in $r^{l+1}$, 
Eq. (\ref{eqZint2}) results in a trivial identity that does not 
determine the arbitrary constants.  Notice that when $\eta=0$ we have 
simply, 
\begin{equation}   
C=-\frac{(\sigma-m\Omega)e^{\nu_c/2}}  
{\left(\sigma-m\Omega+\frac{2m\bar\omega_c}{\Lambda}\right)}  
\end{equation}  
and $D$ is no longer coupled to the other constants, so that, for a 
given $\sigma$, there is only one solution of the equations satisfying 
the conditions at the center. When $\eta\neq 0$, instead, there are 
two independent solutions which satisfy the conditions (\ref{center}) 
at the center. 
  
\subsection{Boundary conditions at the surface}  
As discussed above, in order to integrate the perturbed equations for 
$\eta\neq 0$, an additional condition needs to be imposed at the 
surface of the star. We impose the Israel matching conditions (Israel, 
1966), that is, we require continuity of the induced metric and of the 
extrinsic curvature on the timelike surface which separates the 
interior from the exterior of the star. From the continuity condition 
of the induced metric we learn that $h_0$ is continuous across the 
stellar surface 
\begin{equation}   
[h_{0}]=0\,, 
\end{equation}  
where we use the notation $[f]\equiv \lim_{\epsilon\rightarrow 0} 
\left(f(R_s+\epsilon)-f(R_s-\epsilon)\right).$ Imposing the continuity 
of the extrinsic curvature we find that 
\begin{equation}   
[h_1]=0\,,~~~~[h_0']=0\,.  
\end{equation}  
These conditions are analogous to those found in Appendix B of Price 
\& Thorne, (1969) for the polar case. When applied to our variables, 
they translate in 
\begin{equation}   
[K]=0,\,~~~~[H]=0,\,~~~~[V]=0\,.\label{matching}  
\end{equation}   
Requiring that the interior equations (\ref{eqKint2}--\ref{eqZint2}) 
match with the exterior equations (\ref{eqKext}--\ref{eqHext}) leads 
to the {\it stress-free} condition $\eta X=0$ at the stellar surface, 
which arises directly from equations (\ref{eqHint2}), (\ref{eqHext}). 
  
The physical meaning of the {\it stress-free} condition can be 
understood by looking at the Newtonian case (or at the Cowling 
approximation), in which case one cannot rely on conditions that 
involve metric variables.  Consider the stress acting on an element of 
fluid inside a spherical thin layer bounded by the the surface of the 
star and a sphere of radius $R_s-\epsilon$, and by two lateral 
surfaces defined by $\vartheta_1=$const. and $\vartheta_2=$const.  The 
stress on the upper ($r=R_s$) boundary is zero since there is no fluid 
outside.  If $\epsilon \rightarrow 0$, the stress through the lateral 
surface goes to zero as $\epsilon$ and the only stress is through the 
interior surface.  But, since the volume of the fluid goes to zero 
with $\epsilon$, the stress acting on the interior surface must 
vanish, otherwise the difference would result in an infinite 
acceleration of the thin layer. In the background, the only non 
vanishing radial component of the stress tensor ${\cal S}_{ij}$ 
(i.e. the pressure) vanishes at the stellar surface; the same 
condition must be imposed on the perturbed stress.  Assuming that the 
perturbations of radial velocity and pressure are zero (purely axial 
perturbations), the radial components of the perturbation of the 
stress tensor in spherical coordinates are 
\begin{eqnarray}  
\delta{\cal S}_{rr}=0~, ~~  
\delta{\cal S}_{r \vartheta}=\eta r \frac{\partial}{\partial r} 
\left(\frac{\delta v_\vartheta}{r}\right) ~, ~~  
\delta{\cal S}_{r \varphi}=\eta r\frac{\partial}{\partial r}  
\left(\frac{\delta v_\varphi}{r}\right)\,, 
\end{eqnarray}  
and imposing their vanishing at $r=R_s$, we find  
\begin{eqnarray}  
\eta \frac{\partial}{\partial r} 
\left(\frac{\delta v_\vartheta}{r}\right)_{|r=R_s}=0~, ~~  
\eta \frac{\partial}{\partial r}  
\left(\frac{\delta v_\varphi}{r}\right)_{|r=R_s}=0 \,. 
\end{eqnarray}  
After expanding in spherical harmonics and using eq. (\ref{harmexp}),  
these conditions are equivalent to 
\begin{eqnarray}  
\eta \left(Z' - \frac{2}{r} Z\right) \equiv \eta X = 0\,.\label{newtcond}  
\end{eqnarray}  
This equation shows that, at the boundary between two fluids,
normal shear stresses must be continuous. 
  
In order to better understand the relationship between the Newtonian 
condition (\ref{newtcond}) and the relativistic condition $\eta 
X(R_s)=0$, we note that the shear tensor components 
$\sigma_{r\vartheta}$, $\sigma_{r\varphi}$, expanded in axial vector 
harmonics, give exactly the perturbation function $X$, defined in 
(\ref{Xint}); so our matching condition $\eta X(R_s)=0$ is equivalent 
to the vanishing of $\eta\sigma_{r\vartheta}$, $\eta\sigma_{r\varphi}$ 
at the stellar surface. 
  
On the other hand we recall that the Israel matching conditions, as 
shown in Israel (1966), are equivalent to impose that  
\begin{equation}   
[G_{r\alpha}]=0~~~(\alpha=0,\dots,3)  
\end{equation}  
which, from Einstein's equations, are equivalent to 
$T_{r\alpha}(R_s)=0$. The $\alpha\!=a=\!\vartheta,\varphi$ components 
of this equation give 
\begin{equation}   
T_{ra}(R_s)=\left((\rho+p)u_ru_a+pg_{ra}+\eta\sigma_{ra}  
\right)_{r=R_s}=(\eta\sigma_{ra})_{r=R_s}=0\,, 
\end{equation}  
where we have used the fact that, for a purely axial perturbation, 
$u_r=u^{(0)}_r+\delta u_r=0$ and $pg_{ra}=0$ on the stellar surface. 
  
\subsection{Outgoing wave condition at infinity}  
\label{outgoingcond} 
After imposing the condition $\eta X=0$ at $r=R_s$, for each value of 
the (complex) frequency $\sigma$ the perturbed equations admit only 
one solution extending up to radial infinity.  In practice, we cannot 
extend our integration up to a large distance from the center because 
the amplitude of the outgoing component is exponentially 
increasing\footnote{See Kokkotas \& Schmidt, (1999) for a 
discussion. This problem can be overcome by extending $r$ to the 
complex plane (Andersson {\em et al.}, 1995), or by deriving a 
solution for the equations outside the star in terms of recurrence 
relations (Leaver, 1985; Leins {\em et al.}, 1993).}.  However, we 
found it sufficient to integrate backward, starting from $r=200\,R_s$ 
with an asymptotic expansion of the outgoing wave up to first order in 
$1/r$: 
\begin{eqnarray}  
K^{out}&=&r\left(1+\frac{\alpha}{r}\right)e^{{\rm i}\sigma r_*}  
+O\left(\frac{1}{r}\right)\nonumber\\  
V^{out}&=&{\rm i} r\left(1+\frac{\beta}{r}\right)e^{{\rm i}\sigma r_*}  
+O\left(\frac{1}{r}\right)\label{outgoing}  
\end{eqnarray}  
with  
\begin{eqnarray}  
\alpha&=&\frac{3{\rm i}\sigma-2}{{\rm i}\sigma(1+{\rm i}\sigma)}\nonumber\\  
\beta&=&\frac{2\sigma^2-3}{{\rm i}\sigma(1+{\rm i}\sigma)}\,.  
\end{eqnarray}  
Notice that $K,V$ are proportional to $r$ because $h_0,h_1$, in the 
Regge-Wheeler gauge, are proportional to $r$; however this apparent 
divergence is not a real problem, being only an artifact of the gauge 
choice (see for example Thorne \& Campolattaro, 1967). 
  
The Wronskian $W$ of the two solutions (interior and exterior) of the 
system of equations is 
\begin{equation}   
W=\frac{r-2M}{r^3(\sigma-m\omega)}\left(KV^{out}-VK^{out}\right).  
\end{equation}  
We have numerically checked that $W$ is constant independently of the 
radius at which it is evaluated (typically with 7-8 significant 
digits).

\section{Results}   
\label{results}
We shall now  discuss the results of the numerical integration of 
our equations.
For comparison with previous works and for testing 
purposes, we begin discussing  the case of 
constant density stars,  later  showing the results for
polytropic stars and realistic neutron stars. In each case, we compare the 
Newtonian limit, the Cowling approximation, and the relativistic 
calculation, with the purpose of establishing the qualitative differences 
between the three approaches and to check whether our results converge to known results
in the inviscid limit.

\subsection
{Homogeneous Stars}  
  
For constant density stars the frequencies of the $r-$modes have been shown to lay 
outside the continuous spectrum 
(see e.g. Kokkotas \& Stergioulas, 1999 for the full GR results).  We consider a 
uniform density star with a central energy density of $10^{15}$ 
g/cm$^3$,  mass of 1.086 $M_\odot$ and  radius of 8.02 km 
($M/R=0.2$). We choose the rotational parameter 
$\epsilon=0.3$. In Fig. 1 we show the  $r$--mode 
frequency versus the viscosity parameter $\eta$, assumed to be constant 
throughout the star.  The Newtonian, Cowling, and General Relativistic 
results are shown by dotted, dashed, and solid lines, respectively. 
The Newtonian and GR frequencies in the inviscid limit (1064 Hz and 
1144 Hz) are indicated as dotted and solid  horizontal lines, while the 
shadowed region indicates the continuous spectrum.   Fig. 1 clearly shows 
that as the viscosity  decreases the inviscid Newtonian and GR 
results are recovered, while (as expected) the frequency in the 
Cowling approximation falls inside the continuous spectrum. 
Since the convergence to the inviscid limit  is
reached for $\eta \approx 10^{-5}$ km$^{-1}$, the mode frequency in the 
Cowling approximation can be found by extrapolating the dashed 
line for $\eta \rightarrow 0$; the corresponding value is 1244 Hz, about a ten 
percent larger than the GR value.  It should be stressed that  the mode frequency 
in the Cowling approximation had never been obtained before for perfect fluids.

Since for numerical reasons we must restrict 
our calculation to values of $\eta \gtrsim 10^{-6}$ km$^{-1}$, we are 
always in the region where the viscous damping dominates the 
instability. Therefore,  the inverse imaginary part of the frequency
which we show in  Fig. 2 is a damping, not a growth time. 
Two interesting features are shown in Fig. 2. Firstly, the 
Newtonian, Cowling and GR damping times differ for less than 20\% because 
the metric corrections are small. Notice that in the Newtonian approach
the stellar background is obtained solving the relativistic equations 
and the perturbations are studied using the Newtonian equations of hydrodynamics.
Secondly, all damping times show the  expected 
$1/\eta$ behaviour. More precisely, we 
can compare our results with the analytic back-of-the-envelope formula 
of Cutler \& Lindblom, (1987) for the dissipative time scale of the 
shear viscosity, based on a quasi--uniform density Newtonian model: 
\begin{eqnarray}  
\tau = \frac{\rho R^2}{(l-1)(2l+1)\eta}.   
\label{tau0}  
\end{eqnarray}  
For our model $\tau = 3\times 10^{-8}/\eta$, with $\eta$ 
in km$^{-1}$ and $\tau$ in s.  This estimate is also shown in Fig. 2 
(thin solid line) and it is, surprisingly, in better agreement with 
the GR results than with the Newtonian ones. 
  
It is useful to introduce the Ekman number, which can be defined as 
\begin{eqnarray}  
E_s = \frac{\eta}{2 \rho R^2 \Omega} \approx 10^4  
\left(\frac{\eta}{\rm km^{-1}}\right)  
\left( \frac{P}{\rm ms}\right),  
\end{eqnarray}  
where  $P$ is the period of the mode.
This number represents the ratio between the viscous term and the 
Coriolis force. Typically, it is smaller than $10^{-7}$ (Cutler \& 
Lindblom 1987). For numerical reasons, our calculations cannot be 
extended to values of $E_s$ smaller than 0.1, which is still far from 
real models. Nevertheless, it allows to obtain the $r-$mode 
frequencies in  the inviscid limit with a good accuracy.

\subsection{Polytropic Stars}  
  
We now  turn to polytropic stars. For these models, and  within the same
approximations we use to derive our equations, the $r-$mode was found to disappear for certain 
ranges of the polytropic index $n$ (Ruoff \& Kokkotas, 2001; Ruoff \& 
Kokkotas, 2002).  Only for very compact stars $n<0.8$, the $r-$mode 
frequency lies outside the continuous spectrum and could be found. We 
have considered a polytropic model with $n=1$, and the same 
compactness and rotation parameter as in previous section ($M/R=0.2, 
\epsilon=0.3$). This model has  mass  $M=1.74 M_\odot$ and 
$R=12.86$ km. As shown in  Ruoff \& Kokkotas, 2001, in the 
inviscid case we do not find the $r-$mode because it lies 
inside the continuous spectrum. In Fig. \ref{fig2a} we show the results 
obtained when viscosity is included in the calculation.  In the 
Newtonian case (dotted line), the inviscid limit is nicely recovered 
as before. As for the Cowling (dashed line) and RG (solid line) 
calculations, we can follow the $r-$mode inside the continuous 
spectrum until convergence to the inviscid limit is reached.  The 
corresponding damping times are shown in Fig. \ref{fig2b}. Similarly 
to constant density models, the Relativistic damping time lies between 
the Newtonian and the Cowling calculation, and they agree within a 
factor 2.  For comparison we also include the estimate given by 
Eq. ({\ref{tau0}}) using the average density of the star, which 
overestimates the damping time of about  50 \%.  Notice that using 
the average density in Eq. ({\ref{tau0}}), the damping time (for 
models with constant $\eta$) depends only on $\rho R^2 \propto M/R$, 
therefore it gives the same result for stars with the same 
compactness.  

The results shown in this section can also be compared 
 with a recent work by Villain \& Bonazzola, (2002), 
in which the coupling of the polar and axial perturbations was fully taken into 
account, and   the perturbed Euler equations were integrated in the Cowling 
approximation with spectral methods. For a similar polytropic model 
they found an $r-$mode with a frequency about a 10\% higher than the 
Newtonian value, in good agreement with our results.

To complete our discussion about the existence of $r$-modes in the 
continuous spectrum, in Fig. \ref{fig3} we show the behaviour of the 
real part of the $r-$mode frequency ($\sigma$) as a function of the 
polytropic index $n$ for models with compactness $M/R=0.2$, period of 
1 ms, and a shear viscosity $\eta=10^{-6}$ 
km$^{-1}$. The period and compactness have been chosen to allow for 
direct comparison with the results of Ruoff \& Kokkotas (2001) who could not find 
the $r$-modes  for $n$ larger than a certain 
value, when the real part of the frequency reached the continuous 
spectrum. By introducing a small amount of viscosity, the frequency 
can be calculated for all polytropic indexes, or for any other stellar 
model, even if we stay at the most basic level of approximation: first 
order in the rotation parameter and neglecting the coupling between 
the axial and polar parts. 
Note that when the mode lays outside the continuous spectrum
we obtain results very similar  to Ruoff \& Kokkotas (2001).
 
Some attention must be paid to the fact that viscosity breaks the 
degeneracy of the $r-$modes and allows for the existence of a family 
of modes with an arbitrary number of nodes. A simple analytical 
solution for homogeneous stars in the Newtonian case and for constant 
$\eta$ is detailed in the Appendix.  In this particular case, the 
eigenfunctions are simply given in terms of the spherical Bessel 
functions $Z =  r~ j_l(kr)$. 
For the $l=2$ case we have  
\begin{eqnarray}  
\label{bes}
Z \propto \left( 3 - \frac{1}{(kr)^2} \right) \sin{kr} -  \frac{3}{kr} \cos{kr} \,.  
\end{eqnarray}  
Indeed, we can obtain a family of solutions labeled by different wave-numbers 
$k$.  In Fig. \ref{autof1} we show the first three eigenfunctions of the
Newtonian $r$ modes for constant density stars and for polytropic stars plotted  
with dashed and solid lines, respectively.  The numerical results for 
homogeneous stars have been checked to coincide with the above 
analytical solution. All eigenfunctions have been normalized to their 
surface value. 
The analytical solution shows that, for  low viscosity,  
the leading order correction to the frequency 
is proportional to $\sigma_0 (\eta/\rho)^2 - i \eta/\rho$, with
$\sigma_0=\frac{\Lambda-2}{\Lambda}m\Omega$. 
The linear dependence of the inverse damping time 
with $\eta$ has already been discussed in the previous section; here we notice that 
the correction to the frequency has a quadratic dependence on  $\eta$.
Remarkably, we find that even for 
relativistic perturbations, the numerical results can be well fit by 
this quadratic behaviour with a very  good accuracy. 
  
A last relevant result is that in the fully relativistic calculation we 
could not find more solutions than the fundamental mode. The 
other overtones found to exist in the Newtonian or Cowling approaches, seem 
to be incompatible with the pure outgoing wave condition imposed at 
infinity. In Fig. \ref{autof2} we show a comparison of relativistic 
$r-$mode eigenfunctions for constant density stars (for which the 
inviscid limit is available) with different viscosities: $\eta=0$ 
(solid), $\eta=1.4\times10^{-7}$ km$^{-1}$ (dots), $\eta=10^{-6}$ 
km$^{-1}$ (dashes), and $\eta=10^{-4}$ km$^{-1}$ (dash-dot).  The figure shows that
the eigenfunction of the fundamental $r-$mode 
approaches continuously the inviscid limit, which is close to the $r^3$ dependence.
It should be reminded that both in the Newtonian and in the Cowling approach 
the gravitational perturbations are neglected; thus,
the fact that the relativistic $r-$modes do not admit 
solutions with large wavenumbers may be an indication that they only 
exist for fluids where the axial fluid motion is totally 
decoupled from gravitational perturbations. 

\subsection{Realistic Neutron Stars}  
  
In this section we consider stellar models constructed with realistic 
EOSs and realistic viscosity profiles
with the purpose of establishing if any imprint of the equation 
of state could be carried by the $r-$mode gravitational wave spectrum. 
At low density (below $10^{12}$ g/cm$^3$) we use the BPS (Baym {\em et 
al.}, 1971) equation of state, while for the inner crust $10^{12} 
<\rho < 10^{14}$ g/cm$^3$ we employ the SLy4 EOS (Chabanat {\em et 
al.}, 1998).  At high density ($\rho > 10^{14}$ g/cm$^3$) we will 
consider two different EOSs of neutron star matter representative of 
the two different approaches commonly found in the literature: 
potential models and relativistic field theoretical models. As a 
potential model, we have chosen the EOS of 
Akmal-Pandharipande-Ravenhall (Akmal {\em et al.}, 1998, hereafter 
APR). As an example of the mean field solution to a relativistic 
Walecka--type Lagrangian we have used the parametrization usually 
known as GM3 (Glendenning \& Moszkowski, 1991). For a detailed 
discussion and comparison of many equations of state from the two 
families see the recent review by Steiner {\em et al.} (2005). Setting the 
compactness parameter to $M/R=0.2$, APR  gives a mass of 1.53 
$M_\odot$ and a radius of $11.3$ km, while GM3 gives $M=1.72 
M_\odot$ and $R=12.7$ km.  For a polytropic EOS with $n=1$ the 
corresponding mass and radius for the same compactness are $M=1.74 ~
M_\odot$ and $R=12.8$ km.  
For cold neutron stars below $10^9$ K, neutrons in the inner core become superfluid
and the dominant contribution to the shear viscosity is electron-electron
scattering (see e.g. the review by Andersson \& Kokkotas, 2001); in this regime
$\eta$ can be written as
\begin{eqnarray}
\eta_{ee}=6\times10^{18} \rho_{15}^2 T_9^{-2}~{\rm g/cm/s} = 
1.48\times10^{-15} \rho_{15}^2 T_9^{-2}~{\rm km}^{-1}
\label{etaee}
\end{eqnarray}
with $\rho_{15}$ and $T_9$ being the density and temperature in units of
$10^{15}$ g/cm$^3$ and $10^{9}$ K.
Having this in mind, in this section we will use a viscosity coefficient 
with a quadratic dependence on density
\begin{eqnarray}
\eta= \eta_0 \left(\frac{\rho}{\rho_c}\right)^2 ~{\rm km}^{-1} ~,
\label{eta0}
\end{eqnarray}
where $\rho_c$ is the central density. Since old neutrons stars are
nearly isothermal, we will parametrize our results
as a function of the constant $\eta_0$ that includes the temperature dependence.

In Fig. \ref{fig4a} we show the $r-$mode frequency as a function of $\eta_0$,
comparing the three EOSs: polytrope (dots), APR (dashes) and GM3 
(solid lines). In all cases the rotation period is $P=2$ 
ms. As we can see in the figure, the frequency is rather insensitive 
to the particular details of the EOS, 
provided that the rotation frequency and $M/R$ are the same. 
The Newtonian limit depends only on 
the angular frequency ($\Omega$) and the relativistic correction 
enters through the frame dragging. Since this correction 
goes as $\omega/\Omega \approx I/R^3 \propto M/R$, the leading order 
contribution to the frequency is a function only of the compactness. 

In Fig. \ref{fig4b} we show the damping times for the same models as in Fig. \ref{fig4a}.
For constant density stars, the $\approx \rho R^2$ dependence of the damping 
time translates into a $M/R$ dependence for models with constant 
$\eta$. In realistic, cold ($T\lesssim 10^9$K) neutron stars,
the density is not constant and the viscosity is dominated by the 
electron-electron scattering process (\ref{etaee}). Thus, the 
analytic result (\ref{tau0}) underestimates
the viscous damping time. An improved calculation for $n=1$ polytropes
taking into account the density profiles and using the shear viscosity
(\ref{etaee}) (Andersson \& Kokkotas 2001) gives
\begin{eqnarray}
\tau = 2.2 \times10^{7} \left(\frac{1.4 M_\odot}{M}\right) 
\left(\frac{R}{10{\rm km}}\right)^5  T_9^2~{\rm s}.
\end{eqnarray}
Assuming a general dependence of the viscosity of the form of Eq. (\ref{eta0}),
and using Eq. (\ref{etaee})the above damping time can be shown to satisfy
\begin{eqnarray}
\tau = \left(\frac{1.4 M_\odot}{M}\right) \left(\frac{R}{10{\rm km}}\right)^5
\left(\frac{\rho_c}{10^{15}{\rm g/cm}^3}\right) 
\left( \frac{3.26 \times10^{-8} {\rm km}^{-1}}{\eta_0} \right) ~{\rm s}.
\label{tauak}
\end{eqnarray}
Our relativistic calculations 
show approximately the same dependence on mass and radius as Eq. (\ref{tauak}), but
the damping times are systematically larger of about  60\%. 
We found that a better fit for the realistic
neutron stars (APR, GM3) as well as for the $n=1$ polytrope is given by
\begin{eqnarray}
\tau = \left(\frac{1.4 M_\odot}{M}\right) \left(\frac{R}{10{\rm km}}\right)^5
\left(\frac{\rho_c}{10^{15}{\rm g/cm}^3}\right) 
\left( \frac{5.22 \times10^{-8} {\rm km}^{-1}}{\eta_0} \right) ~{\rm s}.
\label{taugood}
\end{eqnarray}
In Fig. \ref{fig4b} we show, together with the numerical results (thick lines), the results
corresponding to the previous fit (thin lines). The good agreement between them is apparent.

\section{Final remarks and conclusions}   
  
We have investigated the $r-$mode spectrum within a perturbative 
relativistic framework consistently including the effects of shear 
viscosity. A first interesting result of our study is that the continuous spectrum, 
which was an artifact of the level of approximation (first order in 
rotation, neglecting completely the coupling between polar and axial 
perturbations), can be regularized and the frequencies and viscous damping 
times of the modes can be calculated. This can be understood if one 
analyzes the real cause of the existence of the continuous 
spectrum.  By considering only purely axial perturbations, and working 
at first order in the rotation parameter, each spherical fluid layer 
is effectively decoupled from the neighbour (indeed, the perturbations of 
density, pressure, and radial velocity are zero).  Therefore, each 
layer can oscillate independently and have a different characteristic 
frequency. This effect can be produced by differential rotation or, in 
a relativistic framework, by the dragging of the inertial frames. It 
has been discussed in the literature that including the coupling up to 
some level, the shape of the continuous spectrum changes (Ruoff \& 
Kokkotas, 2002), but still one cannot always find a mode. In 
principle, by fully including the coupling (see e.g. Lockitch {\em et 
al.}, 2001; Lockitch {\em et al.}, 2003; Villain \& Bonazzola, 2002), 
one can look for a {\it global mode} that oscillates with a unique 
characteristic frequency, since different layers feel each other and 
their motion is coupled.  We have produced the same effect by 
introducing shear viscosity. The stress between 
the different layers couples them and results in the existence of a 
unique global $r-$mode, instead of a continuous spectrum.  Therefore, 
even working at the most crude level of approximation 
(purely axial perturbations, first order in rotation) one can calculate the 
$r-$modes (complex) frequency for every stellar model. 
  
Another interesting outcome of our study is that, in the 
regime where viscosity dominates, the damping times of relativistic 
modes are about 40-60\% larger than  those obtained in the Newtonian 
case. For homogeneous stars the differences are smaller, and the semi-analytic
estimate obtained by Cutler \& Lindblom (1987) fits well our numerical results.
For realistic neutron stars we find that our relativistic
results give damping times systematically larger (about a 60\%) than
previous Newtonian estimates (Andersson  \& Kokkotas, 2001), but with
the same dependence on mass and radius. An accurate fit to realistic
neutron stars, valid also for $n=1$ polytropes gives
\begin{eqnarray}
\tau = 3.52 \times10^{7} 
\left(\frac{1.4 M_\odot}{M}\right) \left(\frac{R}{10{\rm km}}\right)^5
T_9^2~{\rm s}~.
\end{eqnarray}
Notice that the corrections due to general relativity and the internal
structure almost cancel each other in such a way that the previous
estimate is actually closer to the results for Newtonian constant density
stars (Kokkotas \& Stergioulas, 1999; Andersson  \& Kokkotas, 2001) than
the more elaborated calculation for $n=1$ polytropes (Andersson  \& Kokkotas, 2001).

Therefore, we have shown that by including the 
shear viscosity in the stress-energy tensor we can 
obtain an accurate estimate of the damping time of the $r-$mode in 
neutron stars, provided that realistic equation of state and viscosity 
are known.  We were able to extend our 
results to arbitrarily low viscosity in the Newtonian case, but further 
improvements, such us including higher order couplings
or devising better numerical approaches to deal with smaller values of $\eta$, 
are needed before we can compute accurately 
the window of instability of the $r-$modes in a fully relativistic 
framework. Notice that axial-polar coupling (second order in rotation)
must also be included to take into account the effect of bulk viscosity. 
  
In this paper we have not considered the damping associated to the  
viscous boundary layer in the core-crust interface, which is thought to be  
one of the most efficient mechanisms to damp the $r-$mode oscillations  
(Bildsten \& Ushomirski 2000; Rieutord 2001).
If the crust is assumed to be rigid, the problem  
can be formulated as an Ekman problem in a spherical layer  
(Glampedakis \& Andersson, 2004). In practice this means to  
substitute the boundary conditions at the stellar surface by a  
"no-slip" condition at the core-crust interface, i.e., $Z(r=R_c)=0$,  
and to  include the corrections due to the  
perturbation of the pressure. Shear viscosity in this case is  
essential since it ensures that the perturbations of the velocity  
deviate from the inviscid solution and go to zero in a thin layer near  
the crust (the Ekman layer) with characteristic width $\frac{\Delta  
r}{R} \approx \sqrt{E_s}$. Recent results (Glampedakis \& Andersson,  
2004) show that the damping rate and the correction to the  
frequency due to the presence of the Ekman layer go as  
$(\Omega/R_c)^{1/2}$, with a numerical factor of the order of  
unity. This issue will be addressed in a future work.

\begin{center}  
\bf{Acknowledgements}  
\end{center}  
  
We thank Nils Andersson and Adamantios Stavridis for 
useful comments and discussions. This work has been supported by the 
Spanish MEC grant AYA-2004-08067-C03-02, and the {\it Acci\'on 
Integrada Hispano--Italiana} HI2003-0284.  J.A.P.~is supported by a 
{\it Ram\'on y Cajal} contract from the Spanish MEC.

\begin{center}  
\begin{figure}  
\psfig{figure=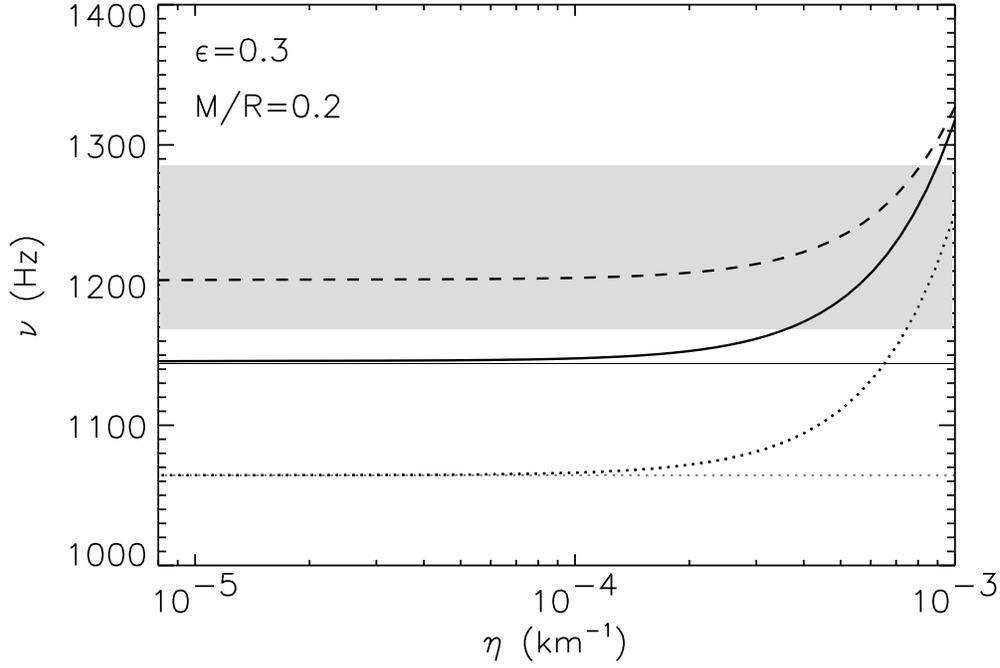,angle=0,width=14cm}  
\caption{The $r-$mode frequency is plotted as a function of the shear viscosity 
parameter $\eta$ for a uniform density model with $M/R=0.2$ and 
rotational parameter $\epsilon=0.3$.  The Newtonian, Cowling and GR 
results are shown, respectively, with dotted, dashed, and solid lines. 
The thin horizontal lines indicate the corresponding Newtonian and GR 
inviscid limits, while the continuous spectrum is the shadowed region.} 
\label{fig1a}  
\end{figure}  
\end{center}  
  
\begin{center}  
\begin{figure}  
\psfig{figure=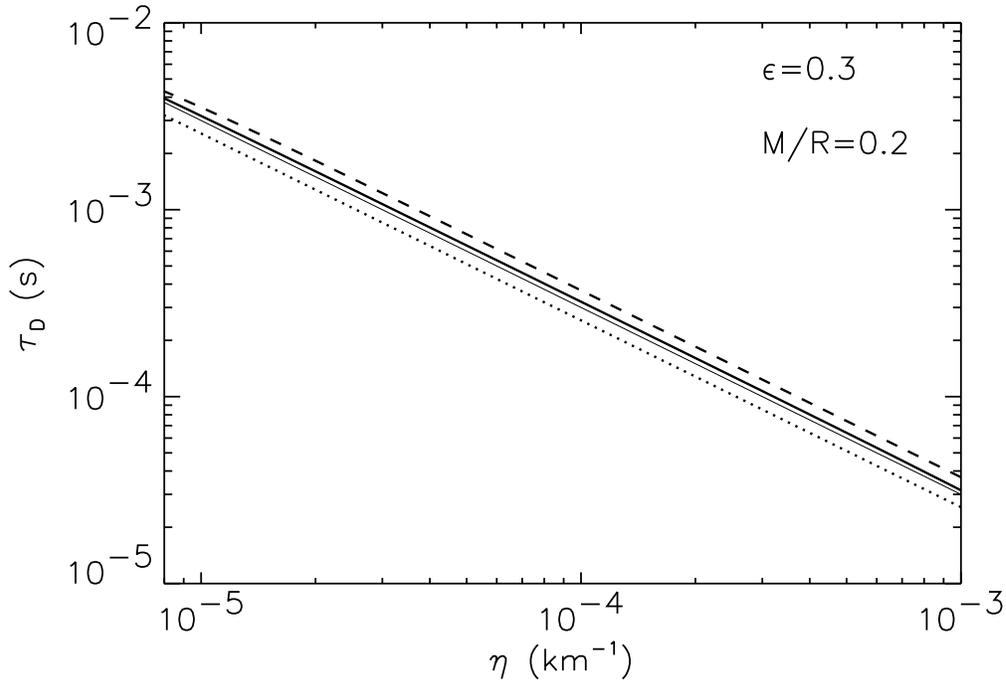,angle=0,width=14cm}  
\caption{The $r-$mode damping time is shown as a function of the shear 
viscosity parameter $\eta$ for the same model as Fig. \ref{fig1a}.
The Newtonian, Cowling and GR results are shown, respectively, with dotted, 
dashed, and solid lines.  The thin solid line refers to the simple estimate obtained using 
Eq. (\ref{tau0}). } 
\label{fig1b}  
\end{figure}  
\end{center}  
  
\begin{center}  
\begin{figure}  
\psfig{figure=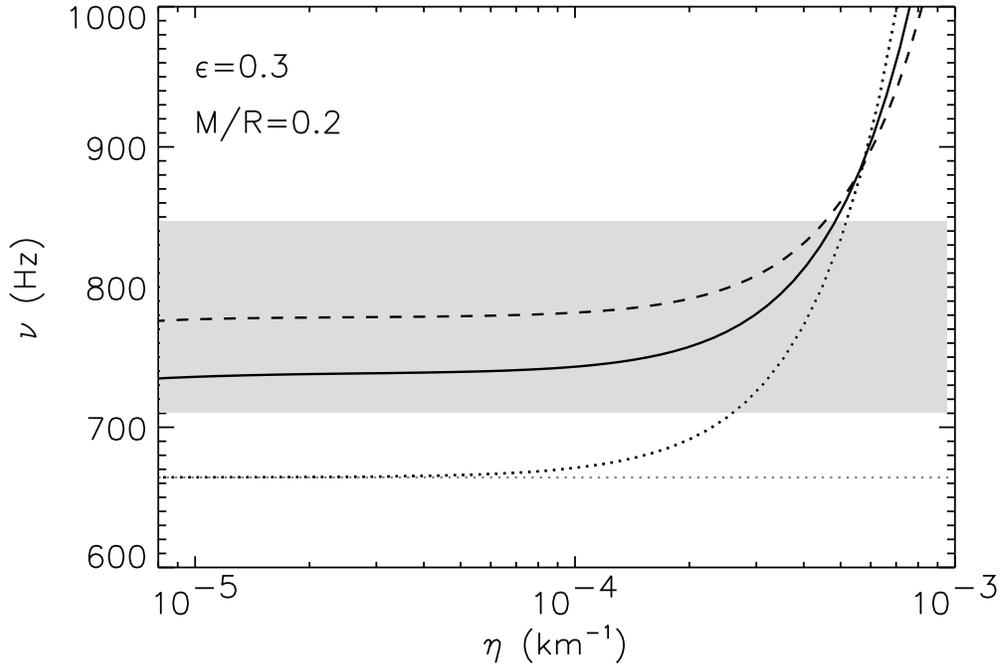,angle=0,width=14cm}  
\caption{The $r-$mode frequency is plotted as a function of 
$\eta$ for a polytropic star with $n=1$, $M/R=0.2$ and 
rotational parameter $\epsilon=0.3$.  The Newtonian, Cowling and GR 
results are shown, respectively, with dotted, dashed, and solid lines. 
The horizontal dotted line indicates the corresponding Newtonian 
inviscid limit, while the continuous spectrum is the shadowed region.} 
\label{fig2a}  
\end{figure}  
\end{center}  
  
\begin{center}  
\begin{figure}  
\psfig{figure=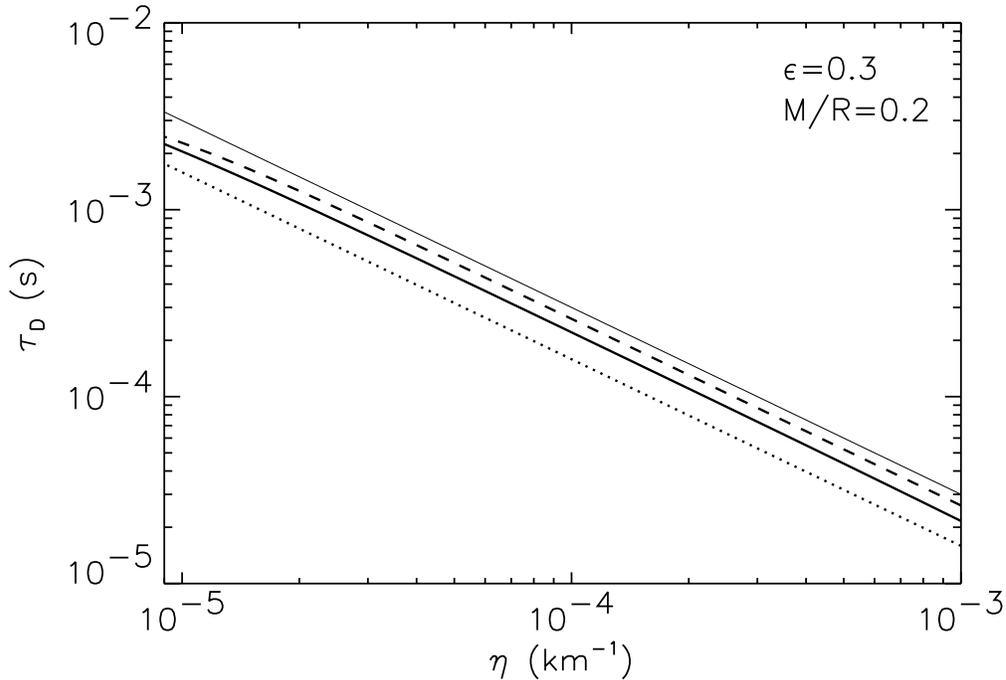,angle=0,width=14cm}  
\caption{The $r-$mode damping time is plotted as a function of 
$\eta$ for the same polytropic star of Fig. \ref{fig2a}.
The Newtonian, Cowling and 
GR results are shown, respectively, with dotted, dashed, and solid 
lines.  The thin solid line refers to the simple estimate obtained using 
Eq. (\ref{tau0}). } 
\label{fig2b}  
\end{figure}  
\end{center}  
  
\begin{center}  
\begin{figure}  
\psfig{figure=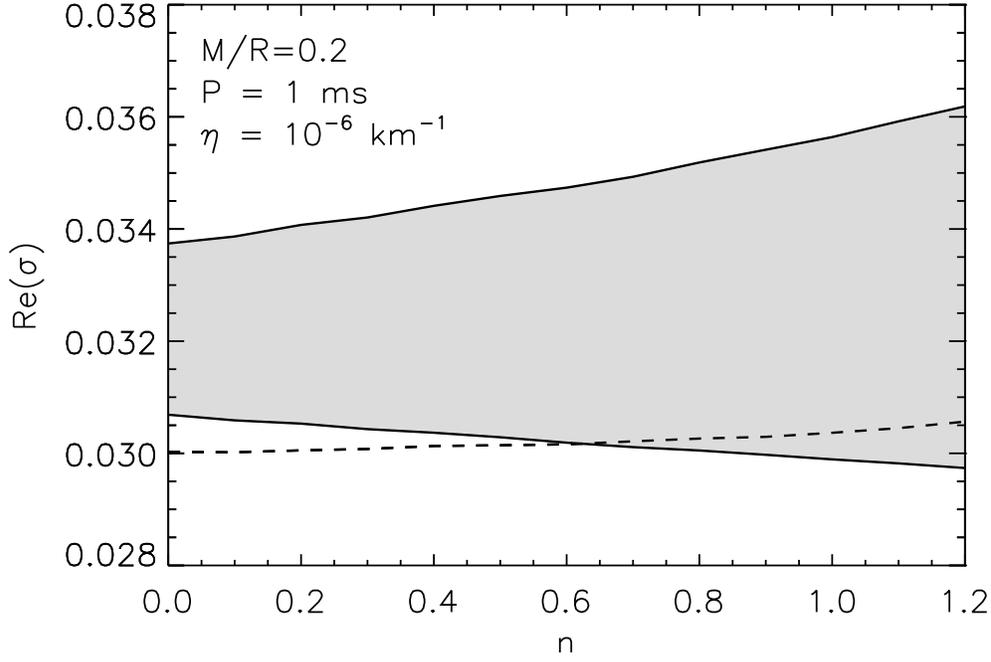,angle=0,width=14cm}  
\caption{The real part of the $r-$mode frequency is plotted
($\sigma$) as a function of the polytropic index $n$ for models with a  
shear viscosity parameter $\eta=10^{-6}$ km$^{-1}$, compactness $M/R=0.2$ and  
period of 1 ms.  }  
\label{fig3}  
\end{figure}  
\end{center}  
  
\begin{center}  
\begin{figure}  
\psfig{figure=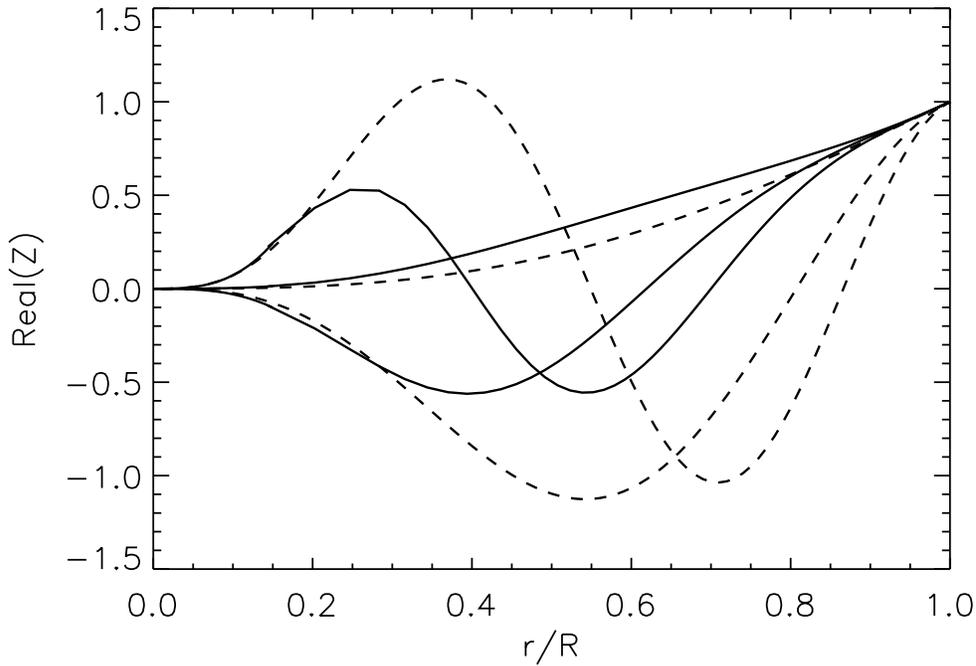,angle=0,width=14cm}  
\caption{Newtonian $r-$mode eigenfunctions for constant density stars 
(dashed lines) and polytropic stars (solid lines) for $\eta=10^{-6}$ 
km$^{-1}$.  The numerical results corresponding to constant density 
stars are found to be in agreement with the analytic eigenfunctions 
derived in the appendix. The eigenfunctions of polytropic stars have 
the same qualitative shape.  } 
\label{autof1}  
\end{figure}  
\end{center}  
  
\begin{center}  
\begin{figure}  
\psfig{figure=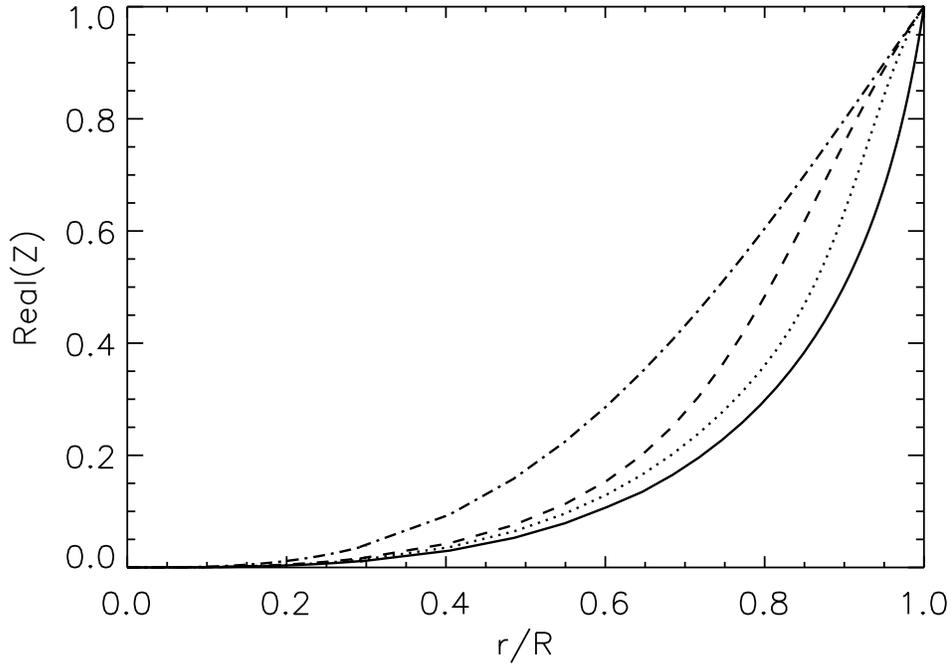,angle=0,width=14cm}  
\caption{Comparison of relativistic $r-$mode eigenfunctions for 
constant density stars and different viscosities: $\eta=0$ (solid), 
$\eta=1.4\times10^{-7}$ km$^{-1}$ (dots), $\eta=10^{-6}$ km$^{-1}$ 
(dashes), and $\eta=10^{-4}$ km$^{-1}$ (dash-dot).  In the 
relativistic case, no mode solution with nodes could be found since
only the fundamental $r-$mode admits a pure outgoing wave solution.} 
\label{autof2}  
\end{figure}  
\end{center}  
  
\begin{center}  
\begin{figure}  
\psfig{figure=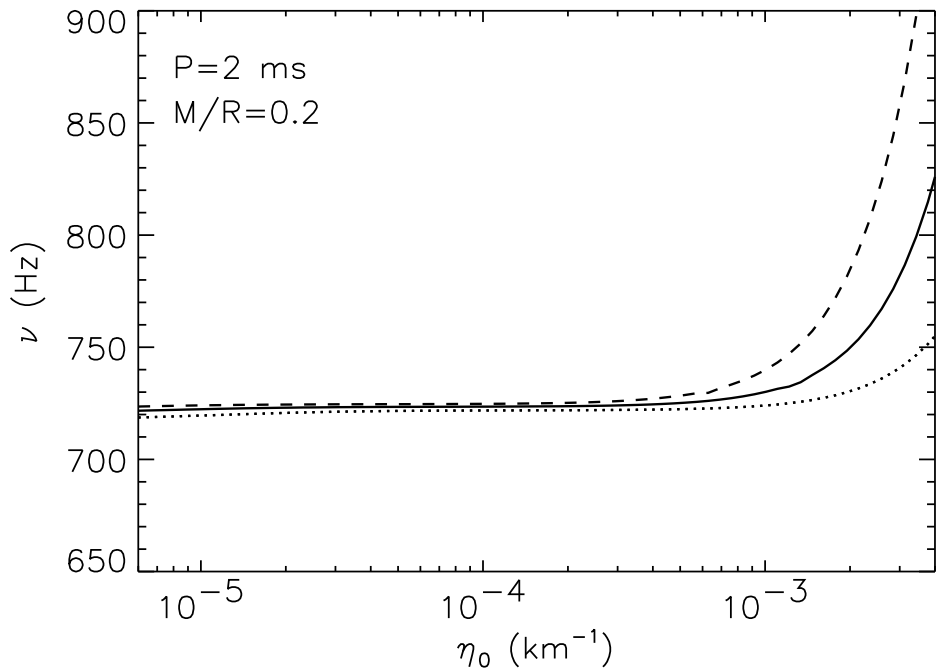,angle=0,width=14cm}  
\caption{Comparison of the $r-$mode frequency as a function of the  
the viscosity coefficient at the center, $\eta_0$, for a polytropic star with $n=1$  
(dotted line), the APR (dashed line) and the GM3 (solid line)  
equations of state. The compactness parameter in all cases is  
$M/R=0.2$ and the rotation period is 2 ms.  }  
\label{fig4a}  
\end{figure}  
\end{center}  
  
\begin{center}  
\begin{figure}  
\psfig{figure=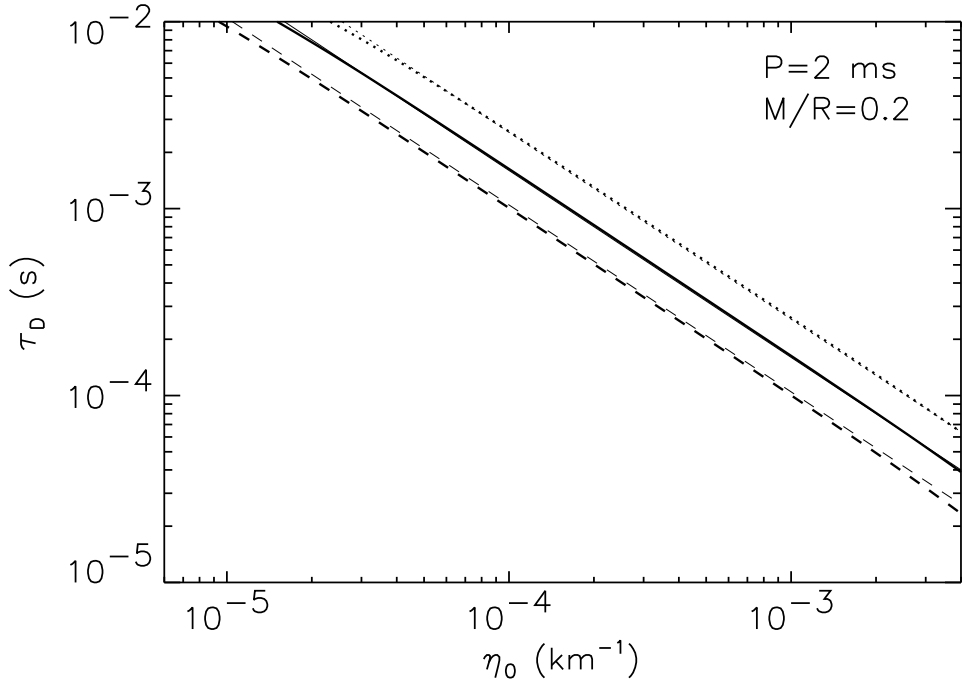,angle=0,width=14cm}  
\caption{Comparison of the $r-$mode damping time as a function of 
$\eta_0$ for a $n=1$ polytrope 
(dotted line), the APR (dashed line) and the GM3 (solid line)  
equations of state, for the same models considered in Fig. \ref{fig4a}.
The thin lines, whoch show the estimates
given by Eq. (\ref{taugood}), are difficult to distinguish from
the real results. }  
\label{fig4b}  
\end{figure}  
\end{center}  
  
\clearpage  
\appendix   
\section{An analytical solution for the Newtonian $r-$modes}  
  
The Newtonian equations (\ref{eqsnewt0}--\ref{eqsnewt}) describing the axial 
perturbations can be written as a single second order ODE 
\begin{eqnarray}  
r^2 Z''-  l(l+1) Z  
+ r^2 \left( \sigma_0 \sigma + i \frac{\rho}{\eta}  
(\sigma-\sigma_0) \right) Z = 0  
\end{eqnarray}  
where $\sigma_0$ is the (real) frequency in the inviscid limit 
$\sigma_0=\frac{\Lambda-2}{\Lambda}m\Omega$.  If we assume that 
$\nu \equiv \frac{\eta}{\rho}$ is a constant, this is just a version of the 
Riccati equation, 
\begin{eqnarray}  
z^2 y'' + \left( k^2 z^2 - l (l+1)\right) y = 0  
\end{eqnarray}  
which has solutions  
\begin{eqnarray}  
y =  kz (A j_l(kz) + B y_l(kz) )  
\end{eqnarray}  
where $j_l(kz)$ and $y_l(kz)$ are spherical Bessel functions of the 
first and second kinds.  In our problem we impose regularity of the 
function at the origin, which implies $B=0$ because $y_l(kz)$ diverge 
in $z=0$. Therefore, we are left with the simple (complex) solution 
\begin{eqnarray}  
Z =  A x j_l(x)  
\end{eqnarray}  
where $x=kz$ and $k^2 = \sigma_0 \sigma + \frac{i}{\nu} 
(\sigma-\sigma_0)$.  Notice that this functions have the correct 
$r^{l+1}$ behaviour near the origin.  The eigenmodes and 
eigenfunctions are then fixed by imposing boundary conditions at a 
given radius $r=R$. Consider the $l=2$ case and let $A=1$, the 
solution is: 
\begin{eqnarray}  
Z(x) = \left( \frac{3}{x^2}-1 \right) \sin{x} -  \frac{3}{x} \cos{x} ~. 
\end{eqnarray}  
The boundary condition $Z'-2Z/r =0$ at $r=R$ implies  
\begin{eqnarray}  
(5 x^2 -12) \sin{x} - x(x^2-12) \cos{x} = 0  
\end{eqnarray}  
which has only real solutions. The first few zeros are approximately $x=2.5, 7.14, 10.5, 
13.8, 17.0 ~~...$.  The frequency of the fundamental $r-$mode 
corresponds to $k R = 2.5$, but there is an infinite number of higher 
order overtones.  For each wavenumber, the deviation from the inviscid 
limit ($s\equiv\sigma - \sigma_0$) can be readily calculated 
\begin{eqnarray}  
s = \frac{\left(k^2 - \sigma_0^2 \right)}{1 + \nu^2 \sigma_0^2}  
(\sigma_0 \nu^2 - i \nu)~.
\end{eqnarray}  
Notice that in the low viscosity limit, the damping time goes exactly 
as $\eta/\rho R^2$, with a multiplicative factor of order unity, but 
the correction to the frequency goes as $\eta^2$.  If the boundary 
conditions are the {\it no-slip} conditions $Z=0$, then we must impose 
that at $r=R_c$ 
\begin{eqnarray}  
\left( 3 -x^2 \right) \sin{x} -  {3x} \cos{x} = 0  
\end{eqnarray}  
whose first zeros are $x= 5.75, 9.1, 12.3, 15.5 ~ ...$  
  
We have checked that the numerical results show the same qualitative 
dependence, even for polytropic or realistic stars. Knowing the 
dependence in the low viscosity limit allows us to regularize the 
equations and solve them numerically with arbitrarily small viscosity. 
Using $\bar{s}=s/\eta$ as a variable, the Newtonian equations 
(\ref{eqsnewt0}--\ref{eqsnewt}) become 
\begin{eqnarray}  
Z'&=& \frac{2}{r} Z + X  
\\  
X'&=& - \frac{2}{r} X +  
\frac{(\Lambda-2)}{r^2} Z  
- (\sigma_0 (\sigma_0 + \eta \bar{s}) + {\rm i}{\rho} \bar{s}) Z  
\end{eqnarray}  
and can be integrated numerically since $\bar{s}$ is well behaved as 
$\eta\rightarrow 0$. 
  
A last important remark is that for low viscosity all the overtones 
have very close frequencies ($\eta^2$ corrections), but the damping 
times scale with $k^2-\sigma_0^2 \approx k^2$, so that modes with shorter wavelength 
are damped faster than the fundamental mode (typically a factor 
4-10). Therefore, understanding the non-linear transfer of energy 
between the fundamental $r-$mode and higher order overtones may be 
relevant for astrophysical applications. 
\clearpage  
  
   
\end{document}